\documentclass[amsfonts,amsmath,prd,preprint,nofootinbib]{revtex4}
\newcommand{\beq}{\begin{equation}}
\newcommand{\eeq}{\end{equation}}

\newcommand{\bsp}{\begin{split}}

\usepackage{epsfig,cancel,ulem}
\graphicspath{{./figures/}}

\begin{document}
\title{\boldmath Compactness, masses and radii  of compact stars within the Eddington-inspired Born-Infeld theory}

\author{A. I. Qauli$^1$}
%\email{ali.ikhsanul@sci.ui.ac.id}
\author{A. Sulaksono$^1$}
%\email{anto.sulaksono@sci.ui.ac.id}
\author{H. S. Ramadhan$^1$}
%\email{hramad@ui.ac.id}
\author{I. Husin$^2$}
%\email{idrushusin@students.itb.ac.id}
\affiliation{$^1$Departemen Fisika, FMIPA, Universitas Indonesia, Depok 16424, Indonesia,}
\email{ali.ikhsanul@sci.ui.ac.id, anto.sulaksono@sci.ui.ac.id, hramad@ui.ac.id}
\affiliation{$^2$Theoretical Physics Lab., THEPI Division, Institut Teknologi Bandung, Jl. Ganesha 10 Bandung 40132, Indonesia.}
\email{idrushusin@students.itb.ac.id}

\def\changenote#1{\footnote{\bf #1}}

\begin{abstract}
We investigate the compactness, masses and radii of realistic neutron stars (NSs) and quark stars (QSs) within the Eddington-inspired Born-Infeld (EiBI) theory of gravity, along with the energy conditions of the corresponding apparent equation of states (EOSs). We show that the maximum compactness and maximum masses constraints extracted from the recent pulsars masses and radii observations can provide the upper and lower limits of $\kappa$ value of EiBI theory. By using BSP parameter set of relativistic mean field (RMF) model to describe NS core EOS including hyperons using standard SU(6) prescription to determine hyperon coupling constants, it can be estimated that $ (2.7 \le \kappa_g \le 7.9) \times$ $\rm 10^{-2} m^5 kg^{-1} s^{-2}$. If we use  confined-isospin-density-dependent-mass (CIDDM) model with additional scalar Coulomb term with  QSK046 parameter set to describe QS EOS, we can obtain lesser value i.e., $ (1.6 \le \kappa_g\le 2.2) \times$ $\rm 10^{-2} m^5 kg^{-1} s^{-2}$. Here, the lower limit of $\kappa$ is obtained by using $M_{\rm max} \gtrsim 2 M_\odot$ constraint of ~Ref.\cite{Compact1} while the upper limit is obtained by compactness $\lesssim$ the maximum compactnes of Ref.~\cite{Compact3}. We have also observed that for large $\kappa$ values, mass-radius relations of NSs and QSs do not exceed causality restrictions because the compactness of NS and QS are saturated after passing certain large critical value. This observation is in agreement with the results obtained in Ref.\cite{Delsate12} for the case of pressure-less stars. We have also found that if  $\kappa$ larger than certain non-zero value, the CIDDM  with vector Coulomb model prediction of QS can reach the maximum mass $\gtrsim$ 2 $M_\odot$. We have found also that these constraint ranges depend significantly on the NS or QS EOS used. If the EOS becomes stiffer, the upper limit of $\kappa$ from compactness constraint becomes smaller and lower limit of  $\kappa$ becomes larger.  We also observe that the uncertainty of the systematic in canonical mass compact stars radii measurements data can affect the $\kappa$ range.  It is also shown that the non-physical apparent EOS of NSs and QSs can satisfy the energy conditions. However, in the NSs case, the square of the sound speed of the corresponding apparent EOSs in the near-surface region is negative. There is also indication that the strong energy condition can be violated by NS and QS EOSs when  $\kappa$ is extremely large. In general, physical requirements for acceptable interior solution for static fluid spheres of GR can be also violated by apparent EOS.
\end{abstract}

\keywords{Apparent EOS, EiBI theory, compact stars}

\pacs{12.39.-x,97.60.Gb,04.50.-h}
\maketitle
\section{Introduction}
\label{sec_intro}
Einstein's general relativity (GR) theory  has  solid conceptual foundation and it passes all precision test for intermediate energy scales with flying colors. However, on cosmological scale GR also faces problems such as  the need of dark energy and dark matter. The theoretical and experimental indications of modification of GR at small and large energies are reviewed in Refs.~\cite{Berti_etal2015,Will2009,Psaltis2008}. In compact stars such as NSs and QSs, gravity plays a relatively dominant role and its collapse leads to large-curvature and strong-gravity environments~\cite{Berti_etal2015,Psaltis2008}. Therefore, the differences in predictions between GR and alternative or modified gravity theories could appear significantly~\cite{DeDeo2003,Kazim2014}. On the other hand,  even though significant  progress has been reported, until now the equation of state (EOS) of a NS is still uncertain (see Refs.~\cite{Lattimer2012,Chamel2013,Lonardoni,Yamamoto,Artyom14,ref:weissenborn,SB2012} and the references therein). Similarly, the QS EOS is also model-dependent~\cite{QS2016,ref:Farhi,ref:isospin,NJL,DSA,ref:cold,ref:hybrid}. It is, however, worthy to note that recently there are some studies about an EOS-independent relation of some NS properties such as moment of inertia, Love number and quadrupole moment as well as the relation of other multiple moments~\cite{YY2013,PA2014,YKPYA2014}.  

Among GR modified theories, the Eddington-inspired Born-Infeld (EiBI) theory attracts quite a lot of attentions recently due to its distinctive features compared to those of GR (see details in Refs.~\cite{JHOR2017, PCD2011,PDC2012,Delsate12} and the references therein).  The theory, proposed by Banados and Ferreira~\cite{Banados10} fusing Palatini approach, is a gravitational analog of a nonlinear theory of electrodynamics known as the Born-Infeld theory~\cite{Born:1934gh,Berti_etal2015,Psaltis2008}. The reviews of the investigations and applications of EiBI theory  can be found in Ref.~\cite{Berti_etal2015} and the references therein. While the most recent comprehensive review of general Born-Infeld inspired modifications of gravity theories and their applications can be found in Ref.~\cite{JHOR2017}. In our context, the EiBI theory is interesting because, it can solve the ``hyperon puzzle'' without introducing new physics in NS matter. 

However, we need to highlight some features of EiBI theory related to this work. The authors of Refs.~\cite{QISR2016,PCD2011,MH2013,Sotani14} have shown that  the Tolman-Oppenheimer-Volkov (TOV) equation version obtained by using EiBI theory could increase or decrease the maximum mass of NS by adjusting the corresponding $\kappa$ value. The author of Ref.~\cite{Sotani14} has also found that through direct observations of the NS radii around 0.5 $M_\odot$ and the precise measurements of neutron skin thickness of $^{208}$Pb, the EiBI theory could be discriminated  from GR. Furthermore, it is also reported that the range of  reasonable values of $\kappa$ parameter in EiBI theory can be constrained by using some astrophysical and cosmological data~\cite{Avelino12}, NSs properties~\cite{QISR2016,PCD2011,PDC2012,Harko13} and the Sun properties~\cite{CPLC2012}. The possibility for distinguishing EiBI theory from GR is also suggested by observing gravitational wave from NS (see Ref.~\cite{Sotani2014_2} for details). Concerning the stellar stability of EiBI theory, the authors of Ref.~\cite{SLL2012} has shown that the standard results of stellar stability still hold in EiBI theory where for a sequence of stars with the same EOS, the fundamental mode $\omega^2$ passes through zero at central density corresponding to the maximum-mass configuration is similar to the one found in GR. Therefore, the corresponding point marks the boundary of the onset of instability where the stellar models with central densities less than the corresponding critical points are stable. Furthermore, The authors of Ref.~\cite{PDC2012} have also shown that there always exists regular solution for compact stars with $\kappa >$ 0 and the corresponding stars have maximum compactness of $\frac{GM}{R} \sim$ 0.3 which is roughly independent from $\kappa$. The collapse constraint, i.e., the compact stars exist if the requirement $\kappa \Delta < 0$ is satisfied, with $\Delta$ is
\begin{equation}
\Delta=(P_c \kappa - 3 \kappa \rho_c -4)(1 + \kappa \rho_c)-\kappa (1-\kappa P_c)(P_c+\rho_c) \frac{ d \rho (P_c)}{d P_c},
\nonumber
\end{equation}
here $P_c$ and $\rho_c$ are the central pressure and density of the stars. It means that if the EOS is thermodynamically consistent, the onset of the star stability region in EiBI theory depends only on $P_c$ and $\kappa$. We also need to note that the EiBI theory shows also a singularity associated with the phase transition matter for negative $\kappa$ due to the appearance of discontinuity in energy density around the transition region~\cite{Sham2013}. The curvature singularities appearing at surface of compact stars within EiBI theory for polytropic EOS have been already discussed for examples in Refs.~\cite{BSM2008,PS2012,PSV2013,Kim2014}. The key issue of these singularities are that higher-order derivatives of matter fields, which appear in the EiBI field equations as a results of integrating out nondynamical degrees of freedom, make the geometry sensitive to sharp in matter configuration. Wheter these singularity rule out the theory or due to an artifact of the fluid approximation etc is still under debate (see Ref.~\cite{PS2012} and the references therein).

In the case when the matter of  the star can be described by an ideal fluid with barotropic EOS, the modified field equations within EiBI with physical EOS (The ideal fluid energy-momentum tensor is derived from the microscopic interaction among the constituents of the corresponding matter) in respect to auxiliary metric $q$ can exactly  be expressed as field equations of GR but with effective ideal fluid energy-momentum tensor. The corresponding EOS has highly nonlinear relation between the pressure and the energy density. This EOS is called the apparent EOS in Ref.~\cite{Delsate12}. The observations like NSs as well as other compact object properties are difficult to distinguish this degeneracy~\cite{Delsate12,Berti_etal2015}. Note that even the EOS of matter in flat space time satisfies all energy conditions, but the apparent EOS could violate strong energy condition (SEC) or real $\tau$ condition of EiBI theory. The corresponding violation is demonstrated in Ref.~\cite{Delsate12} in the case of dust (pressure-less) EOS. The important matter properties in GR encoded in various energy conditions which have some physical consequences (see details in Ref.~\cite{Delsate12} and references therein) and furthermore, the EOS is also restricted by some requirements in order the interior solution for static fluids spheres of GR to be physically meaningful~\cite{MH2013}. This violation can be understood because  the apparent EOS is defined in $q$ metric. Therefore, it is indeed not a physical EOS. Here, we also  study the apparent EOS of actual compact stars such NS and QS. This study could complement the previous results~\cite{Delsate12}.

In this work, we investigate the role of the compactness extracted from the most recent masses and radii of the existed pulsars analysis obtained by the authors of Ref.~\cite{Compact3} as well as the recent masses\cite{Compact1,Compact2,Rezzolla:2017aly} and 1.4 $M_\odot$ radii\cite{Steiner:2017vmg,Steiner:2015aea,Nattila:2015jra,Guillot:2014lla,Annala:2017llu} of compact stars observations to constraint $\kappa$. These observations are used to update the the lower limit as well as to obtain the upper limit of $\kappa$  of NSs and QSs. We also explore the effect of the stiffness of the corresponding EOSs to the allowed $\kappa$ range obtained. For completeness, we also report the compatibility of the NS and QS apparent EOSs with energy conditions  up to a quite wide range of $\kappa$ values.

The paper is organized as follows. Sec.\ref{sec_Formalism}, describes briefly the formalisms used in this work. Sec.~\ref{sec_RD} is devoted to discuss the results while  the conclusion is given in Sec.~\ref{sec_conclu}.

\begin{figure}[t]
\centering
\includegraphics[width=.6\linewidth]{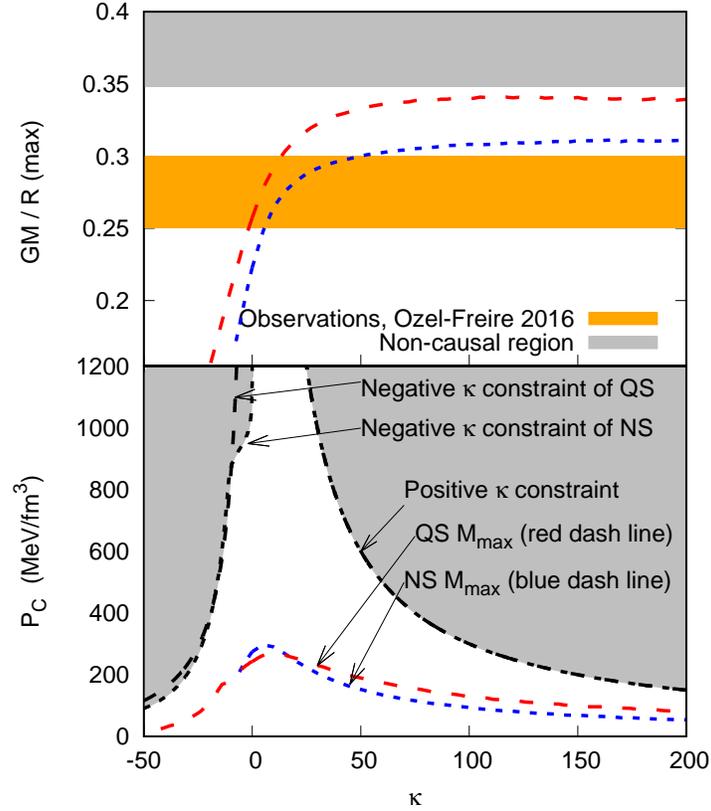}
\caption{ Maximum compactness as a function of $\kappa$ (upper panel), and central pressure as a function of $\kappa$ (lower panel). On the upper panel, non-causal region due to maximum compactness $\simeq 0.35$ is taken from Ref.\cite{Causality}, while the constraint from observation data is taken from masses and radii data extracted from recent data of the properties of pulsars analysis~\cite{Compact3}.  In lower panel, the black dot and black-dash dot plots are generated based on the requirement that the value of $\tau$ must be real and the corresponding constraint apply for both NS and QS, while the QS and NS $M_{max }$ plots are the $P_c$ and $\kappa$ values which predict the corresponding maximum masses. NS EOS calculated by using RMF BSP parameter set, while QS EOS calculated by using CIDDM with additional scalar Coulomb term  with $\kappa_3$=2500,  $\kappa_2^{(1)}$=0.3 and  $\kappa_1$=0.46 (QSK046).  }
\label{fig:gmr_kappa}
\end{figure}

\begin{figure}[t]
\centering
\includegraphics[width=.62\linewidth]{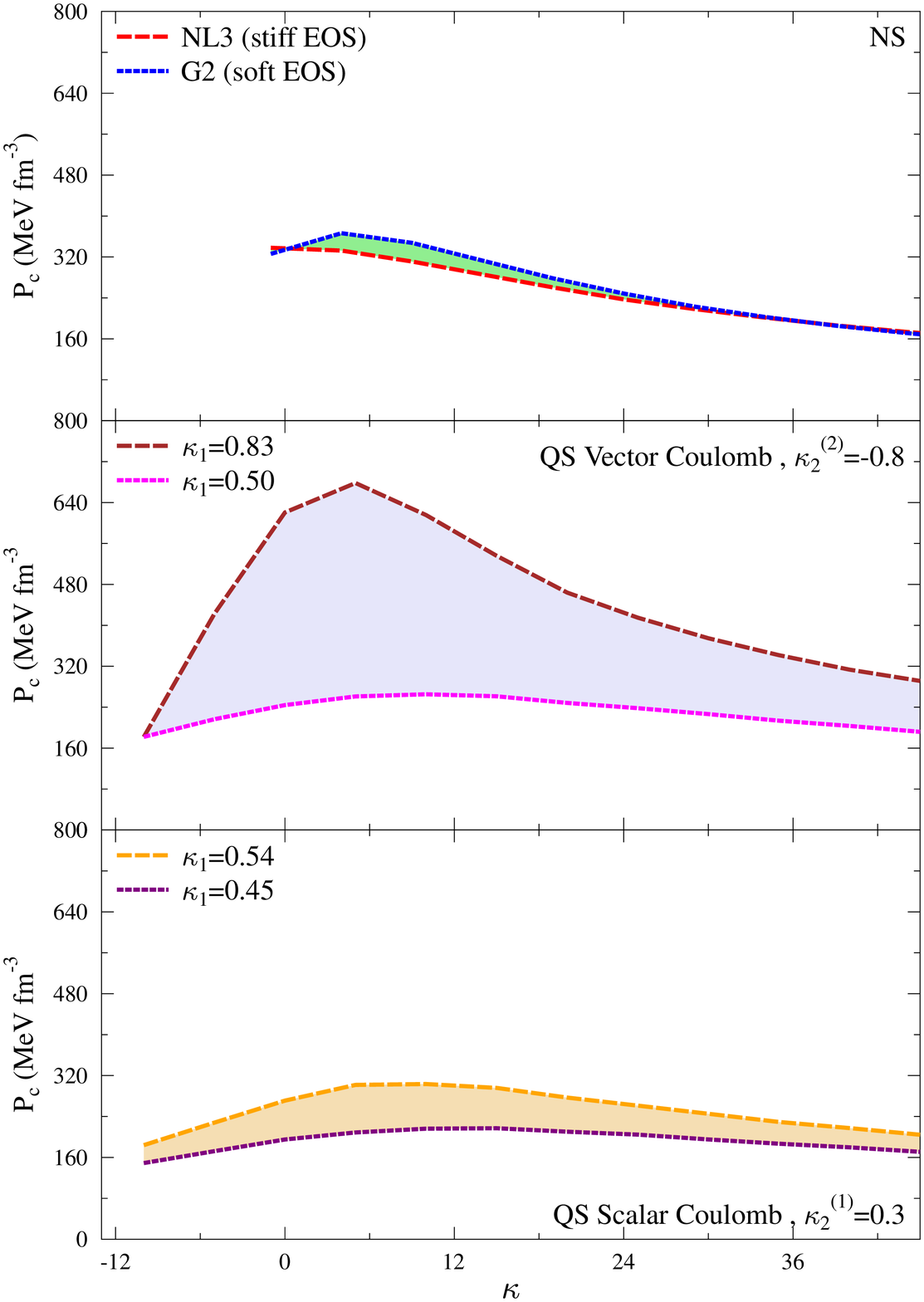}
\caption{ Central pressure as a function of $\kappa$   for stiff which is represented by NL3 parameter set and soft which is represented by G2 parameter set NS EOSs (upper panel), similar to the one in upper panel but for QS within CIDDM model with vector Coulomb (middle panel) and for  QS within CIDDM model with scalar Coulomb (lower panel).}
\label{fig:PCkappa}
\end{figure}

\begin{figure}[t]
\centering
\includegraphics[width=.62\linewidth]{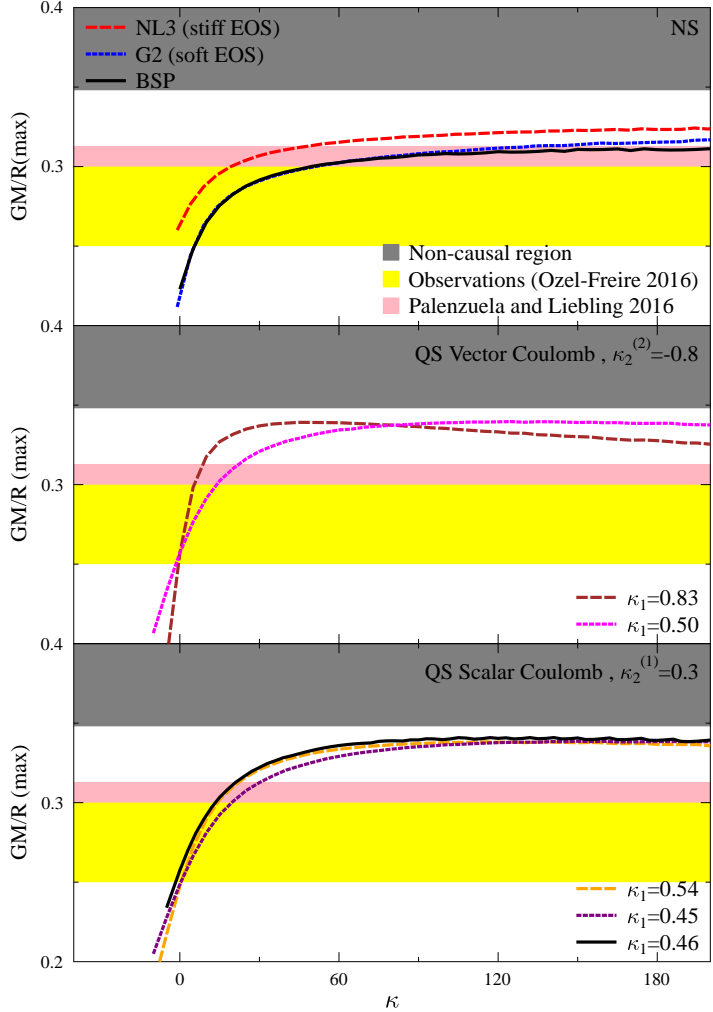}
\caption{ Maximum compactness as a function of $\kappa$ for stiff and soft  NS EOSs are represented by NL3 and G2 parameter sets, respectively (upper panel) while for QS within CIDDM model with vector Coulomb (middle panel) and the ones for QS within CIDDM model with scalar Coulomb (lower panel). Note that we use the same constraints as Fig.~\ref{fig:gmr_kappa} plus additional constraint from Pelenzuela and Liebling~\cite{Compact4} for comparison.  }
\label{fig:EOSCompactness}
\end{figure}

\begin{figure}[t]
\centering
\includegraphics[width=.6\linewidth]{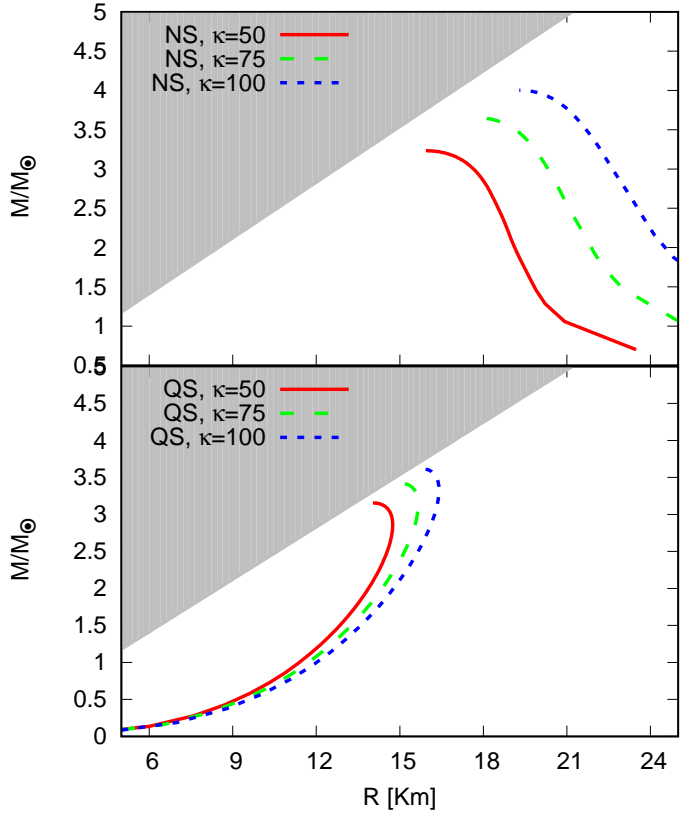}
\caption{Mass-radius relation of NSs (upper panel) and QSs (lower panel) for large $\kappa$ values. The gray-shaded area is non-causal region deduced from the critical compactness $\frac{G M}{R} \simeq 0.35$ \cite{Causality}. Each line represents the mass-radius relation for a particular value of $\kappa$.}
\label{fig:radmass}
\end{figure}

\begin{figure}[t]
\centering
\includegraphics[width=.62\linewidth]{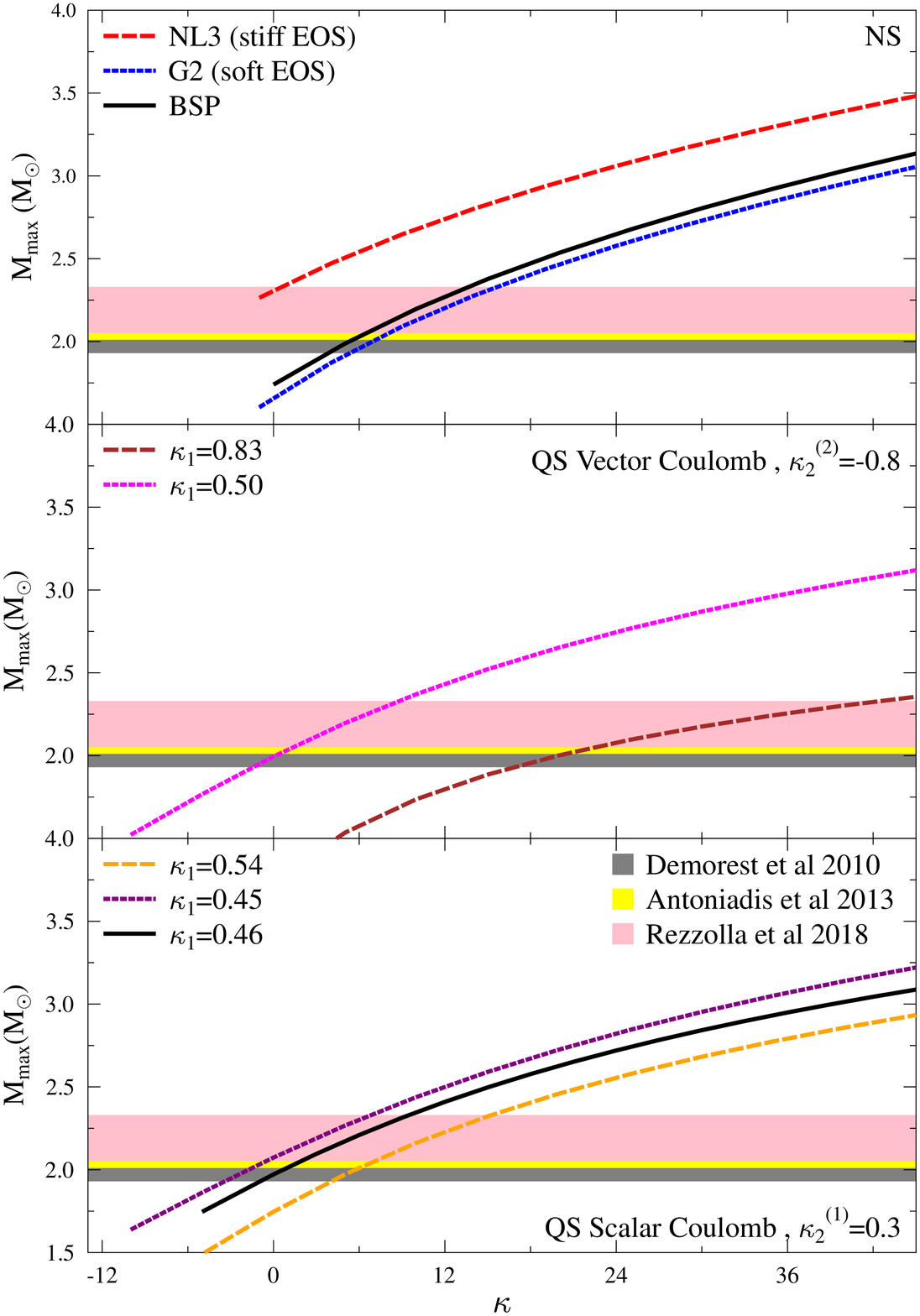}
\caption{ Maximum mass as a function of $\kappa$   for stiff which is represented by NL3 parameter set and soft which is represented by G2 parameter set of NS EOSs (upper panel), similar to the one in upper panel but for QS within CIDDM model with vector Coulomb (middle panel) and for  QS within CIDDM model with scalar Coulomb (lower panel). We use the maximum mass constraints from Refs.\cite{Compact1,Compact2,Rezzolla:2017aly}.}
\label{fig:MaxMasses}
\end{figure}

\begin{figure}[t]
\centering
\includegraphics[width=.62\linewidth]{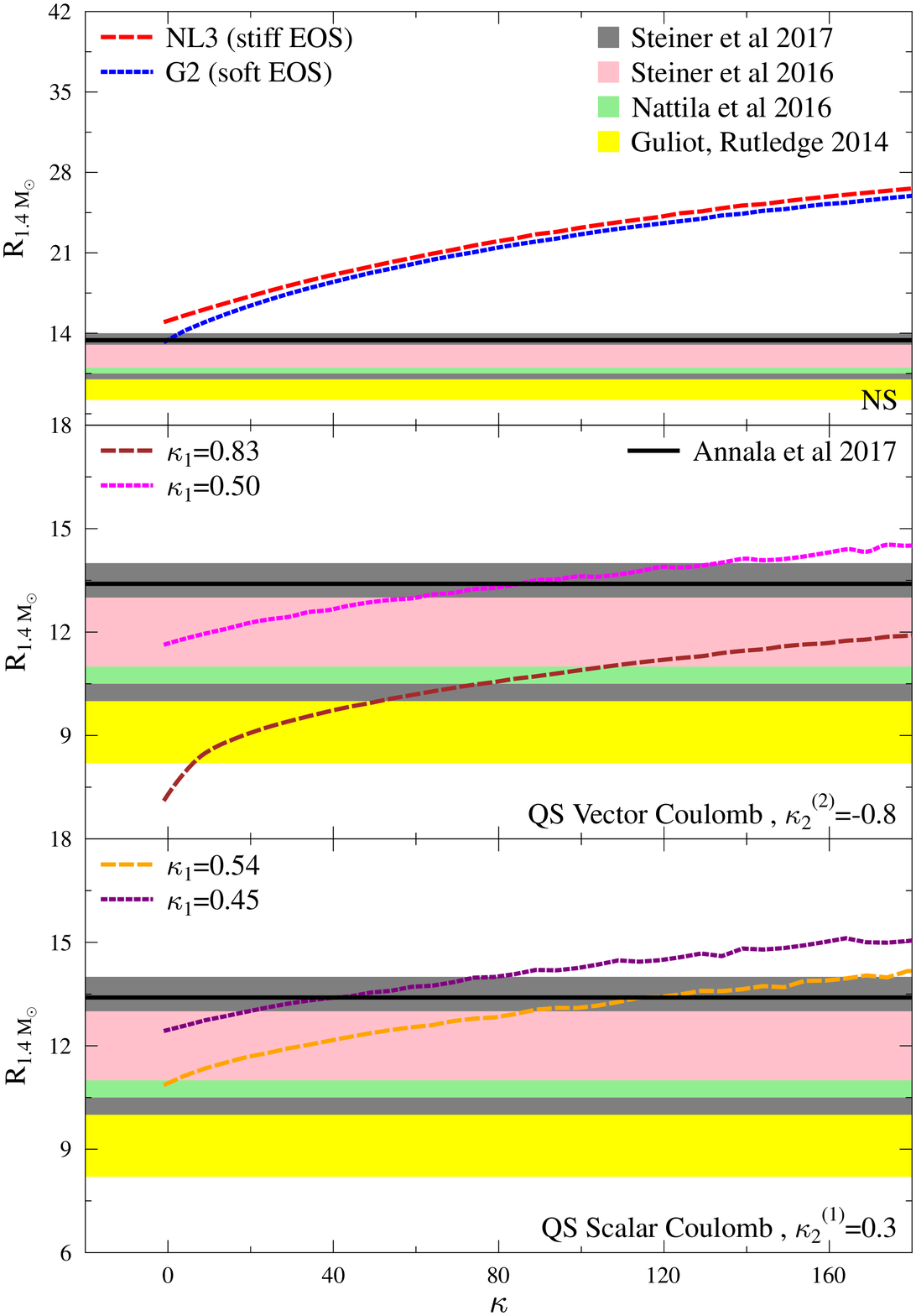}
\caption{ Radius range of 1.4 $M_{\odot}$ as a function of $\kappa$   for stiff which is represented by NL3 parameter set and soft which is represented by G2 parameter set of NS EOSs (upper panel), similar to the one in upper panel but for QS within CIDDM model with vector Coulomb (middle panel) and for  QS within CIDDM model with scalar Coulomb (lower panel). We use the radius constraints from Refs.\cite{Steiner:2017vmg,Steiner:2015aea,Nattila:2015jra,Guillot:2014lla,Annala:2017llu}}
\label{fig:1.4MSunradii}
\end{figure}

\section{Formalisms}

In this section, we shall briefly discuss the EOS of matter and the corresponding compact star constraints used as well as review the EiBI theory formalism, by focusing more on the coupling of compact stars matter with gravity.  Note that throughout this paper we use units c = G = 1, and in addition the parameter $\kappa$ in this unit is ($\rm 10^6 m^2$) unit while in standard SI unit it becomes $\kappa_g$=$8~\pi~G~\kappa$ in $\rm m^5 kg^{-1} s^{-2}$ unit.

\label{sec_Formalism}
\begin{figure}[t]
\centering
\includegraphics[width=.9\linewidth]{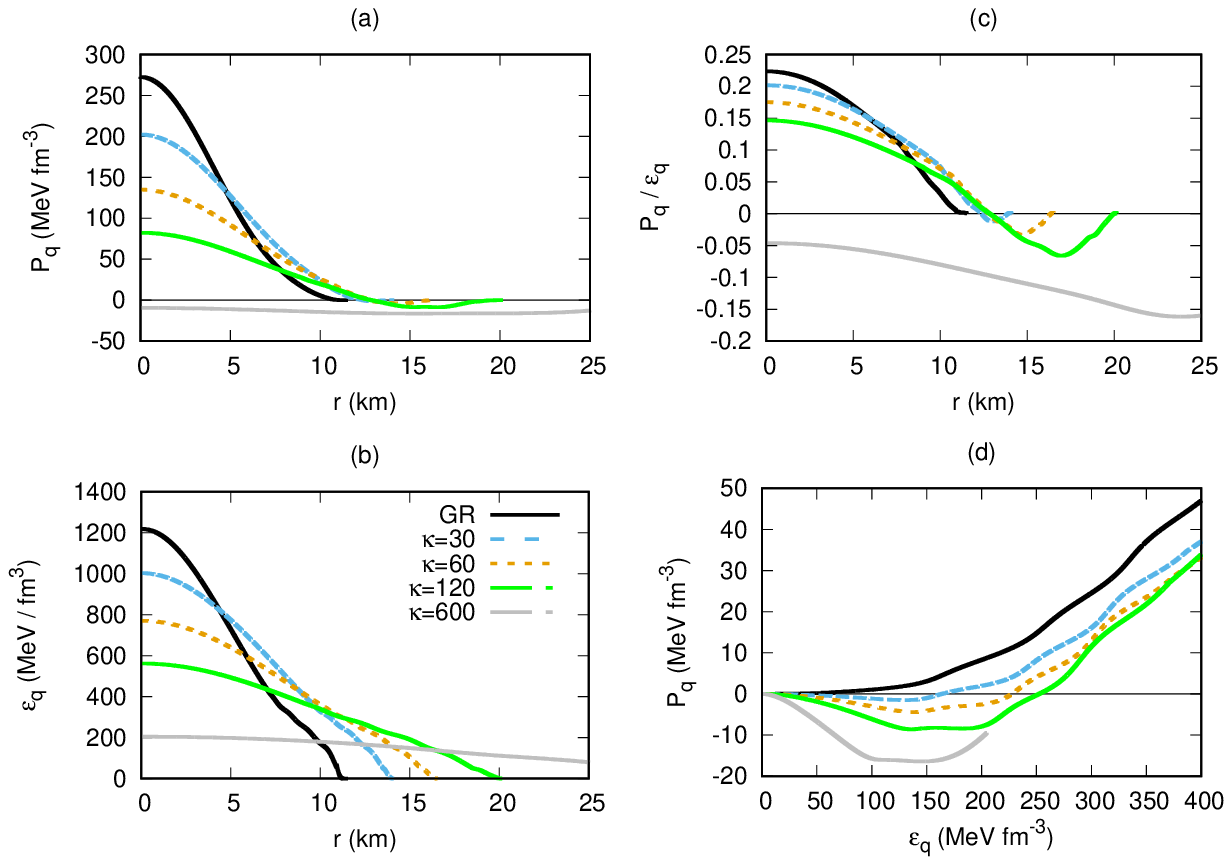}
\caption{The profile for NSs with maximum mass. (a) The left-upper panel is the apparent pressure, (b) left-down panel is the apparent energy density, (c) right-upper panel is the ratio of pressure to energy density, and (d) right-down panel is the apparent EOS. Note that the apparent properties used in the plots are deducted from properties of maximum mass of NS for each $\kappa$ value, as in Fig~[\ref{fig:gmr_kappa}].}
\label{fig:NS}
\end{figure}

\subsection{Neutron and quark stars equation of states}

In this subsection, we briefly discuss the EOS of the compact stars used in this work. The EOS of the NS core is calculated by using the relativistic mean field model with BSP parameter set~\cite{SB2012} under which the standard SU(6) prescription and hyperon potential depths~\cite{JSB_AG} are used  to determine the hyperon coupling constants. For the inner crust EOS, we used polytropic EOS while for the outer crust region of the neutron star, the EOS of R\"uster {\it et~al.} \cite{Ruester:2005fm}  is employed. Note indeed that the masses and the radius of NS or QS strongly depend on the adopted EOS. However, not all RMF parameter sets predictions are simultaneously compatible with the ones of pure neutron matter EOS at low density predicted by fundamental theory, experimental data of finite nuclei global properties, heavy ion experimental data predictions for symmetric nuclear matter and pure neutron matter EOSs at sub-saturation densities, etc. The BSP parameter set is one of the RMF parameter sets whose predictions pass all the corresponding tests (for example, see Refs.~\cite{SB2012,QISR2016} and references therein for more details about the features of EOS used in this work). Therefore, we use the BSP in this work as a representative parameter set with acceptable EOSs in quite a wide density range. To see the dependency of the constrained range of $\kappa$ on EOS we used also the EOS predicted by RMF NL3 parameter set as a stiffer EOS and RMF G2 parameter set as s softer EOS compared to BSP (see Ref.~\cite{SB2012} and the references therein for detailed of the corresponding parameter sets). We need also to note that in many works, if we consider that the hyperons should be present in order that we can obtain 2 $M_\odot$ predictions\cite{Compact1,Compact2} in GR, we should modify the standard prescriptions to determine the hyperons coupling constants or add new physics in EOS (see Refs~\cite{ref:weissenborn,SB2012,JSB_AG} and the references therein). 

For QS we used the EOS based on the confined-isospin-density-dependent-mass (CIDDM) model with additional scalar Coulomb term of strange quark matter with parameter $\kappa_3$=2500,  $\kappa_2^{(1)}$=0.3 and  $\kappa_1$=0.46. We call this parameter set QSK046 for short. Here $\kappa_3$ is isospin dependent parameter,  $\kappa_2^{(1)}$ is scalar Coulomb parameter and $\kappa_1$ is harmonic oscillator parameter. This EOS also passes the test for acceptable EOS to describe the quark matter (see the details of the EOS in Ref.~\cite{QS2016} and the references therein). Note that the authors of Ref.~\cite{QS2016} have found that within GR, if the Coulomb term is included, for the models where their parameters are consistent with SQM absolute stability condition, the 2 $M_\odot$  constraint prefers the maximum QS mass prediction of the model with the scalar Coulomb term to that of the model with the vector Coulomb term. To see the dependency of the constraining $\kappa$ on EOS we used also the EOSs predicted the range of allowed QS EOS  of CIDDM model by SQM stability condition not only for scalar Coulomb but also for vector Coulomb (See Ref.~\cite{QS2016} for detail discussions of the role of Coulomb terms in the CIDDM model). Note that for vector Coulomb  we employ $\kappa_3$=2500,  $\kappa_2^{(2)}$=-0.8 and $0.5 \le \kappa_1 \le 0.83$ while for scalar Coulomb we use  $\kappa_3$=2500,  $\kappa_2^{(2)}$=0.3 and $0.45 \le \kappa_1 \le 0.54$. 

\subsection{Compactness, maximum mass and radii constraints}
Here we briefly discuss the compact stars constraints used in this work. For compactness constraint we used  the compactness range calculated by using the data of recent radius and mass of pulsars systematically analysis extracted from many different observation results obtained by the authors of Ref.\cite{Compact3}. For comparison, the compactness range obtained\cite{Compact3} is not significantly different from the range extracted by using the combination only of recent masses and radii of pulsars from Refs.\cite{Compact1,Compact2,Causality}  which was used in Ref.\cite{Compact4} to constraint the parameter $\beta$ in scalar-tensor theory. The causality limit taken from Ref~\cite{Causality} is used also as an additional compactness constraint.

The NS maximum mass establishment comes from the result of two accurate NS mass measurements. The mass 1.97 $\pm$ 0.04 $M_\odot$ of pulsar J1614-2230 is measured from the Shapiro delay~\cite {Compact1} and the mass 2.01 $\pm$ 0.04 $M_\odot$~\cite{Compact2} of pulsar J0348+0432 is measured from the gravitational redshift optical lines of its  white dwarf companion. In addition, there are evidences that some black widow pulsars might have higher masses. For example, pulsar B1957+20 reportedly has a mass of $M_G$ = 2.4 $\pm$ 0.12 $M_\odot$~\cite{vanKerkwijk:2010mt}, and even gamma-ray black widow pulsar J1311-3430~\cite{Romani:2012rh} has higher mass than  B1957+20 but with less accuracy. Recently, Rezzolla  {\it et~al.}~\cite{Rezzolla:2017aly} reported by combining the gravitation wave observations of merging systems of binary neutron stars and quasi-universal relations between the maximum mass of non-rotating stellar model, they can set the range of limit maximum NS mass as  $2.01^{+0.04}_{-0.04}\le M_{\rm TOV}/M_{\odot} \le 2.16^{+0.17}_{-0.15}$. This NS maximum mass limit poses a tighter constraint on the equation of state (EOS) of dense matter in the NS core. Note that in Ref.~\cite{QISR2016}, the author used the range of NS maximum mass to constraint the lower bound of $\kappa$ of NS. Here we also investigate also the effect of the recent limit range of NS maximum mass obtained in~\cite{Rezzolla:2017aly} to update the lower bound of $\kappa$ of NS.

The accurate measurements of the NS radii would also tightly constrain the properties of the matter in NS core.  Unfortunately, however, the analysis methods used to extract NS radii from observational data still have high uncertainty and mostly they come from systematics and the limits of recent observational radii from different sources or even from the same source are often in contradictory one to another (see the detail discussions for examples in Refs.\cite{Steiner:2017vmg,Compact3,Sulaksono:2014wna} and the references therein).  

By analyzing eight quiescent low-mass X-ray binaries in globular clusters, the authors of Ref.~\cite{Steiner:2017vmg} have found that the radius of a 1.4$M_\odot$ NS is most likely from (10-14) km and the tighter constraints are only possible with stronger assumptions about the nature of NS, systematics of the observations and the nature of dense matter. Other works suggest stronger constraints than the one of Ref.~\cite{Steiner:2017vmg} for example Ref. ~\cite{Steiner:2015aea} obtained canonical NS radii between 11 and 13 km by assuming the chiral effective theory approaches to neutron matter can be employed above the nuclear saturation density. However, the uncertainty of the corresponding assumption is difficult to fully quantify. Other examples are the results of Gulliot and Rutledge~\cite{Guillot:2014lla} with quite small canonical NS radii (9.4 $\pm$ 1.2) km and N\"attil\"a {\it et~al.} ~\cite{Nattila:2015jra} have found the corresponding radius in the range of (10.5 $\le$ $R_{1.4 M_{\odot}}$ $\le$ 12.8 )km. Recent analysis by combining the LIGO/Virgo detection of gravitational waves originating from a neutron-star merger GW170817 and the existence of two-solar-mass NS predict the maximal radius of  a 1.4$M_\odot$ NS is 13.5 km\cite{Annala:2017llu}.  We need also to note that Suleimanov {\it et~al.} have found that the lower limit on the NS radius 14 km for masses below 2.2$M_\odot$, independently of the chemical composition~\cite{Suleimanov}. In this work, we will compare the radius of 1.4$M_{\odot}$ of all NS and QS EOS with the observational ones obtained by the authors of Refs.~\cite{Steiner:2017vmg,Steiner:2015aea,Nattila:2015jra,Guillot:2014lla,Annala:2017llu}. We discuss also the dependency of the uncertainty of allowed $\kappa$ by the uncertainty of the observation 1.4$M_\odot$ radii data.

\begin{figure}[t]
\centering
\includegraphics[width=.9\linewidth]{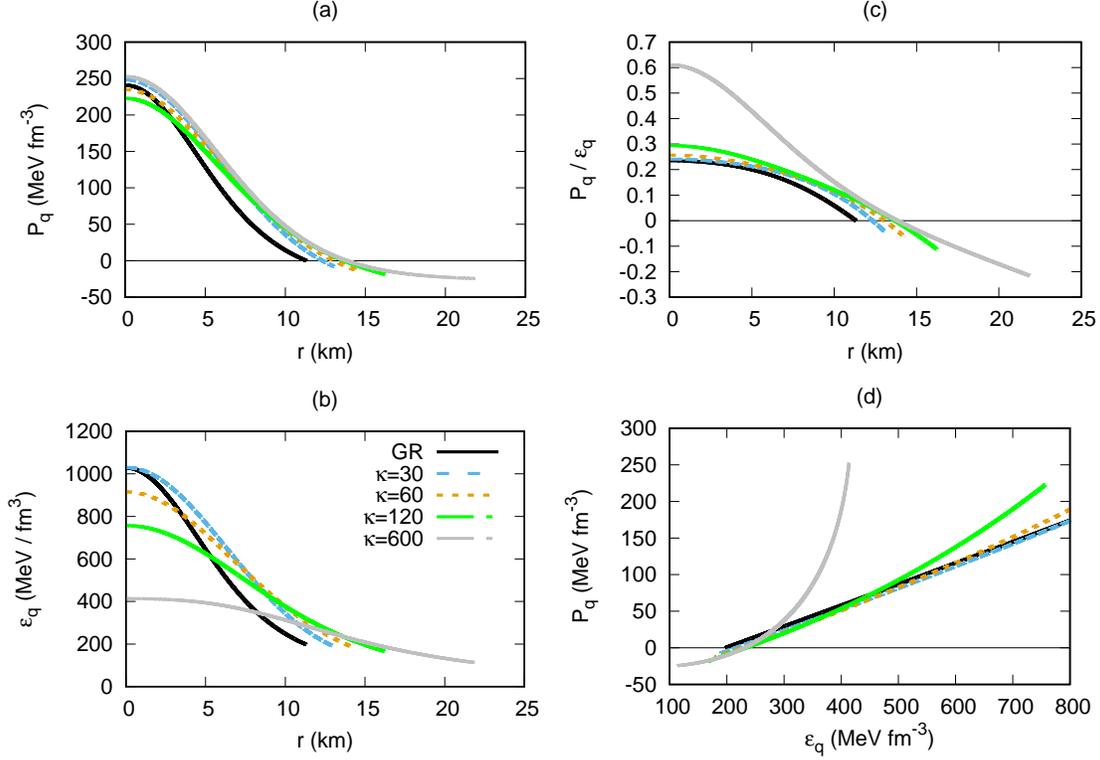}
\caption{The profile for QSs with maximum mass. (a) The left-upper panel is the apparent pressure, (b) left-down panel is the apparent energy density, (c) right-upper panel is the ratio of pressure to energy density, and (d) right-down panel is the apparent EOS. Note that the apparent properties used in the plots are deducted from properties of maximum mass of QS for each $\kappa$ value, as in Fig~[\ref{fig:gmr_kappa}].}
\label{fig:QS}
\end{figure}

\subsection{EiBI theory as GR with apparent EOS in respect to auxiliary  metric $\bf q$}

The action of  EiBI theory of gravity given by~\cite{Banados10,Harko13,Delsate12}
\begin{eqnarray}
\nonumber  S &=&\frac{1}{8\pi\kappa} \int d^4x \left({\sqrt{-|g_{\mu\nu}+\kappa R_{\mu\nu}|} - \lambda \sqrt{-g}}\right)\\
&& + S_M[g,\Psi_M], \label{a}
\end{eqnarray}
where  $R_{\mu\nu}$ is the symmetric Ricci tensor. In the Palatini formalism $R_{\mu\nu}$ is a functional of  connection $\Gamma^\alpha_{\mu\nu}$, $R[\Gamma]$, and the connection and the (physical) metric $g_{\mu\nu}$ are treated as two independent fields. Meanwhile, $\kappa$ and  $\lambda$ are parameters related to the Born-Infeld non-linearity and the cosmological constant, respectively. The action (\ref{a}) reduces to the ordinary Einstein-Hilbert when $\kappa\rightarrow0$. Here $|g_{\mu\nu} + \kappa R_{\mu\nu}|$ denotes the absolute value of the determinant of the tensor $(g_{\mu\nu} + \kappa R_{\mu\nu})$.

Varying the action (\ref{a}) with respect to $\Gamma$ and $g$, we obtain the following equations: 
\begin{eqnarray}
\lambda q_{\mu\nu} &=& g_{\mu\nu} + \kappa R_{\mu\nu}, \label{b}\\
q^{\mu\nu} &=& \tau \left(g^{\mu\nu} - 8 \pi \kappa  T^{\mu\nu} \right), \label{c}\\
\Gamma^\alpha_{\beta\gamma} &=& \frac{1}{2} q^{\alpha\rho}(q_{\rho \beta,\gamma} + q_{\rho \gamma,\beta} - q_{ \beta \gamma,\rho}), \label{d}
\end{eqnarray}
where $q_{\mu\nu}$ is the so-called ``auxiliary" metric, $\tau \equiv \sqrt{g/q}$, and $q$ is the determinant of metric $q_{\mu\nu}$. 

From Eqs.(\ref{b}) and (\ref{c}), one can find the mixed Einstein tensor $G^\mu_\nu$ for $q_{\mu\nu}$ as (See Ref.\cite{Delsate12} for details),
\begin{eqnarray}
\nonumber G^\mu_\nu[q_{\mu\nu}] &\equiv& R^\mu_\nu - \frac{1}{2} R \delta^\mu_\nu = 8\pi\mathcal{T}^\mu_\nu - \Lambda \delta^\mu_\nu, \label{e}
\end{eqnarray}
 with
\begin{eqnarray}
\mathcal{T}^\mu_\nu\equiv\tau T^\mu_\nu +\left(\frac{\tau - 1 - 4 \pi \tau \kappa T}{8 \pi \kappa}\right)\delta^\mu_\nu,
\end{eqnarray}
and the cosmological constant $\Lambda= (\lambda-1)/\kappa$. The function $\tau$ can be obtained by multiplying Eq. (\ref{c}) by metric $g_{\nu\alpha}$ and then taking its determinant~\cite{Delsate12}
\begin{eqnarray}
\tau = {|\left({\delta^\mu_\nu - 8 \pi \kappa T^\mu_\nu}\right)|}^{-\frac{1}{2}}.
\end{eqnarray}
Because it is assumed that the cosmological constant do not significantly influence the compact star properties, henceforth, we set $\lambda \equiv1$. 

The standard EOS model for compact stars is that the energy-momentum tensor assumes the form of perfect fluid, i.e.,
\begin{eqnarray}
T_{\mu\nu} = (\epsilon + p)u_\mu u_\nu + p g_{\mu\nu},\label{tmn}
\end{eqnarray}
which satisfies the conservation equation, $\nabla_\mu T^{\mu\nu}=0$. In Eq.~(\ref{tmn}), $\epsilon$, $p$, and $u_\mu$ denote the actual energy density, the isotropic pressure, and the four-velocity of the NS matter, respectively. It is shown in Ref.~\cite{Delsate12} that under this assumption, it is possible to re-express $\mathcal{T}^\mu_\nu$ in the form of perfect fluid in terms of  $q_{\mu\alpha}$, instead of $g_{\mu\alpha}$, with an apparent fluid velocity ${v}^\mu$ obeys ${v}^\mu{v}^\alpha q_{\mu\alpha}$=-1, while the apparent pressure $P_q$ and energy density $\epsilon_q$ become
\begin{eqnarray} 
P_q &=& \tau P+\mathcal{P}\nonumber\\ 
\epsilon_q&=&\tau \epsilon-\mathcal{P},
\end{eqnarray}
with
\begin{eqnarray} 
\mathcal{P} &\equiv& \frac{\tau - 1 - 4 \pi \tau \kappa (3 P-\epsilon)}{8 \pi \kappa},\nonumber\\
\tau &=& {[(1+8 \pi \kappa \epsilon)(1-8 \pi \kappa P)^3]}^{-\frac{1}{2}}.\label{tau}
\end{eqnarray}
It is obvious from Eq.~(\ref{tau}) that in the limit of $\kappa \rightarrow 0$, then $\mathcal{P} \rightarrow 0$ and $\tau\rightarrow 1$.  From this view, EiBI becomes GR with additional isotropic gravitational pressure $\mathcal{P}$ in apparent stress tensor $\mathcal{T}^\mu_\nu$ in respect to $q$ metric. Here $\tau$  should be real number and it depends on the actual EOS.  The information of the difference between the apparent and actual EOS encodes in  $\tau$ and $\mathcal{P}$. 

It is known that the acceptable EOS should satisfy the energy conditions. The corresponding energy conditions are \cite{Delsate12}:
\begin{itemize}
\item null energy condition (NEC)
\begin{equation}
\epsilon + P \geq 0,
\label{nec}
\end{equation}
\item weak energy condition (WEC)
\begin{equation}
\epsilon + P \geq 0, ~{\rm and} ~\epsilon \geq 0,
\end{equation}
\item strong energy condition (SEC) 
\begin{equation}
\epsilon + P \geq 0, ~{\rm and} ~\epsilon + 3P \geq 0,  
\end{equation}
\item dominant energy condition (DEC) and causal energy condition (CEC) 
\begin{equation}
\epsilon \geq |P|, ~{\rm and} ~|\epsilon| \geq |P|.
\label{cec}
\end{equation}
\end{itemize}
We need also to point out that in order to be physically meaningful, the interior solution for static fluid spheres of GR must also satisfy some general physical requirements, such as (See Ref.~\cite{MH2013} and the references therein for details):
\begin{itemize} 
\item the density $\epsilon$ and pressure $p$ should be positive inside the star\item the gradients $\frac{d \epsilon}{dr}$ and $\frac{dp}{dr}$ should be negative,
\item inside the static configuration the speed of sound should be less than the speed of light, 
\item the interior metric should be joined continuously with the exterior Schwarzschild metric,
\item the pressure $p$ must vanish at the boundary $r$ = $R$ of the sphere.
 \end{itemize}

As mentioned in introduction, if we consider EiBI theory as GR with an apparent EOS, then for completeness, it is interesting also to check  whether the apparent EOS  of compact stars obey also those requirements or not. Note that because the   apparent EOS is not physical or actual EOS, the violation of these requirements are not an indication that the theory has a problem.

\subsection{Compact stars in EiBI theory}

Here we provide the TOV equations of EiBI theory version. The line element of the physical ($g_{\mu\nu}$) and the auxiliary ($q_{\mu\nu}$) metrics that describe the structure of compact static and spherically symmetric objects \cite{Sham2013,Harko13} are
\begin{eqnarray}
g_{\mu\nu}dx^\mu dx^\nu = -e^{\nu(r)} c^2dt^2 + e^{\lambda(r)}dr^2 + f(r) d\Omega^2,\nonumber\\
q_{\mu\nu}dx^\mu dx^\nu = -e^{\beta(r)} c^2dt^2 + e^{\alpha(r)}dr^2 + r^2 d\Omega^2.
\end{eqnarray}
By using these definition for functions $a$ and $b$ as
\begin{eqnarray}
a \equiv \sqrt{1+8 \pi \kappa \epsilon },\label{eq:ab1}\\
b \equiv \sqrt{1-8 \pi \kappa P },
\label{eq:ab2}
\end{eqnarray}
we can obtain~\cite{Harko13,QISR2016} 
\begin{eqnarray}
m' = \frac{1}{4 \kappa} \left({2 + \frac{a}{b^3} - \frac{3}{ab}}\right)r^2,
\label{TOVa}
\end{eqnarray}
and a similar form as the one of GR, pressure derivative, can also be obtained as  
\begin{eqnarray}
P' = -\frac{b}{4 \pi  \kappa}\frac{ab(a^2 - b^2)\left(\frac{1}{2\kappa}(\frac{1}{ab}+\frac{a}{b^3}-2)r^3 + 2 m\right)}{r^2\left({1-2 m}\right)[4ab^2 + (3a - bc_q^2)(a^2 - b^2)]},\nonumber\\
\label{TOV}
\end{eqnarray}
where $c_q^2 = \left(\frac{da(b)}{db}\right)$ =$-\frac{b}{a}\left(\frac{d\epsilon}{dp}\right)$ and the prime in $p$ and $m$ in Eqs. ~(\ref{TOVa}) and (\ref{TOV}) means the first derivative of the corresponding variables in respect to $r$. We need to note that EiBI and GR theories are identical for the region outside the star (r $\ge$ $R$). Therefore, we can use the same boundary conditions at $r=R$ as those of GR for solving TOV equations. In the end, we can obtain static properties of compact stars based on the EiBI theory of gravity by explicitly solving Eqs.~(\ref{TOVa}) and (\ref{TOV}) using the corresponding EOS of NS~\cite{QISR2016} or QS~\cite{QS2016} as an input.

\section{Results and discussion}
\label{sec_RD}

In the lower panel of Fig.~\ref{fig:gmr_kappa}, we show the allowed region of $P_c$ and $\kappa$ within the EiBI theory for NS represented by with BSP parameter set and QS with QSK046, respectively. The right and left gray-shaded regions are the excluded areas by the requirement that $\tau\in\Re$. For positive $\kappa$, the right gray-shaded region is excluded by  the $P_c \le \frac{1}{8 \pi \kappa}$ constraint, and for negative $\kappa$ the left gray-shaded region is excluded by   the $\epsilon_c \ge \frac{1}{8 \pi \kappa}$ constraint. The difference of NS and QS for negative $\kappa$ value close to 0 is due to the different composition of both stars, so that for the same $P_c$ both stars have different $\epsilon_c$. These results are quite in-line with the ones reported in Ref. \cite{Delsate12}. Each point in blue dot and red dash lines are obtained from the value of $P_c$ and $\kappa$ of NS and QS, with maximum masses, respectively. The region below these lines are the onset of stability regions of the NSs and QSs based on the EiBI theory. It is clear that the onset of stability regions of realistic model of the stars (NS with BSP and QS with QSK046 parameter sets) exist inside the acceptable $\kappa$ region from the one of the real $\tau$ constraint. It also is interesting to observe that the stability condition for compact stars within EiBI constrains the maximum $P_c$  instead of $\kappa$ i.e., at $P_c\approx$ 300 $\rm MeV~fm^{-3}$ and  $\kappa\approx$ 5 ($\kappa_g\approx$ 0.0083) for NS and $P_c\approx$ 270 $\rm MeV~fm^{-3}$ and  $\kappa\approx$ 12 ($\kappa_g\approx$ 0.020) for QS. The  onset of stability of NS and QS is quite compatible with the one obtained from the relation $\kappa \Delta < 0$~\cite{PDC2012}. At certain negative value of $\kappa$ i.e., $\kappa\approx$ -8 ($\kappa_g\approx$ -0.013), the NS becomes unstable. This can be seen in the lower panel of Fig.~\ref{fig:gmr_kappa} where the stability boundary line (blue-dashed line) of NS cannot be continued after reaching the negative value of $\kappa$. For the case of QS the value of the corresponding negative $\kappa$ is significantly lower i.e., $\kappa\approx$ -50 ($\kappa_g\approx$ -0.084). This could be due to the role of dilute EOS crust in NS. The reason of the instability has been already discussed in Refs.~\cite{Kim2014,Sham2013}.

The dependency of $P_c$ vs $\kappa$ relation on the EOSs used are shown in Fig.~\ref{fig:PCkappa} where the upper panel for the case NS predicted by RMF model where the stiff EOS is represented by NL3 parameter set and soft EOS is represented by G2 parameter set. Middle panel for the case QS within CIDDM model using vector Coulomb term with the range of parameter is allowed by stability condition of strange quark matter while the lower panel is similar to the one in middle panel but for CIDDM model using scalar Coulomb term. It can be seen that the $P_c$ vs $\kappa$ relation for NS and QS with scalar Coulomb cases do not significantly depend on the stiffness of the EOS used, while for QS with vector Coulomb quite depends on the stiffness of the corresponding EOS. However, even do no show explicitly, in overall, all NS and QS EOSs are still compatible with negative dan positive $\kappa$ constraints in Fig.~\ref{fig:gmr_kappa}.  
        
In the upper panel of  Fig.~\ref{fig:gmr_kappa}, we show the maximum compactness as a function of $\kappa$. The gray-shaded area in the figure is the constraint form the region which is excluded by causality where the value of the compactness should be $\lesssim 0.35$ \cite{Causality}. For the constraint from observation, we take compactness value extracted from the data of three pulsars masses and radii analysis from Ref.~\cite{Compact3}. It is expressed in the figure by the yellow-shaded area.  It can be observed that the maximum compactness of the NS and QS stars predicted by represented parameter sets (BSP for NS and  QSK046 for QS) saturate at large $\kappa$ and they do not reach accusal region. It means that the EiBI stars have maximum compactness which is roughly independent on $\kappa$. This result is consistent with the one obtained by Ref.\cite{PDC2012}. However, It can be observed also that the limit of  maximum compactness of QS is closer to the  non-causal region compared to the one of NS.  The compactness constraint from the observational data of Refs.~\cite{Compact3} can be used as restricted constraints for the upper limit of $\kappa$. Here we obtain $\kappa \lesssim13$ ($\kappa_g\lesssim$ 0.022) for QS and $\kappa \lesssim47$ ($\kappa_g\lesssim$ 0.079) for NS. Note that Ref.~\cite{QISR2016} used 2 $M_{\odot}$ to obtain lower limit of $\kappa$ for NS i.e., 4 $\lesssim$ $\kappa$ $\lesssim$ 6 or 0.006 $\lesssim$ $\kappa_g$ $\lesssim0.01$. Refs.~\cite{PCD2011,Avelino12} obtained $|\kappa_g|\lesssim0.01$ for NS. It is reported in~\cite{CPLC2012} that the sun properties can be used to constrain $\kappa$, that  $|\kappa_g|\lesssim3\times 10^5$. This constraint is significantly looser than the ones of NS upper limit $\kappa_g$. If the saturation of energy density $\epsilon_0=2.68 \times 10^{14} \rm g ~cm^{-2}$ the authors of Ref.\cite{PCD2011} reported that moment of inertia of NS expected from the future observation can be estimated about $|8 \pi \kappa_g \epsilon_0|\lesssim0.1$. It was also shown in Ref.~\cite{Sotani14} that one could distinguish EiBI with $8 \pi \kappa_g \epsilon_0$ $\lesssim0.03$ from GR from fundamental frequency of stellar oscillation, independent of the EOS for NS matter.

In Fig.~\ref{fig:EOSCompactness}, we show the impact of stiffness of the NS and QS EOSs on the range of $\kappa$ constraining by maximum compactness. Maximum compactness as a function of $\kappa$ for stiff NS EOS which is represented by NL3 parameter set and soft one which is represented by G2 parameter set are shown in upper panel. For QS within CIDDM model with vector Coulomb is shown in middle panel and for  QS within CIDDM model with scalar Coulomb is shown in lower panel. Note that we used the same constraints as Fig.~\ref{fig:gmr_kappa} and with additional larger compactness maximum constraint from Pelenzuela and Liebling~\cite{Compact4}. It is obvious if the NS or QS EOS is stiffer, the upper limit of $\kappa$ becomes smaller and if we use larger compactness constraint\cite{Compact4}, the upper limit of $\kappa$ is larger.  For example, if we use NL3 parameter set which is stiffer than the ones of BSP or G2 parameter set, we obtain upper limit with $\kappa \lesssim$ 19 ($\kappa_g \lesssim$ 0.031). This value is smallar than those of BSP and G2 parameter sets. If we use constraint from Ref.~\cite{Compact4}, we obtain  slightly larger upper limit of $\kappa$  compared the one obtained by using constraint from  Ref.~\cite{Compact3}, i.e., $\kappa \lesssim$ 48 ($\kappa_g \lesssim$ 0.081). Note that the maximum compactness extracted from Ref.~\cite{Compact3} use larger data of pulsars masses and radii than those of the one used in Ref.~\cite{Compact4}. This the reason that we choose to use the one of Ref.~\cite{Compact3} as the main maximum compactness constraint in Fig.~\ref{fig:gmr_kappa}. It can be seen also in Fig.~\ref{fig:EOSCompactness} that all EOSs used do not pass the non-causal region. The result in the upper panel of Fig.~\ref{fig:gmr_kappa} is also consistent with the mass-radius relations shown in Fig.~\ref{fig:radmass}. It can be seen in the upper panel for NS and lower panel for QS of Fig.~\ref{fig:gmr_kappa} that as the value of $\kappa$ increases the maximum mass and the corresponding radius of both stars do also increase. It causes the compactness to increase up to a certain point at high $\kappa$ value, then it saturates. Therefore, the causality restriction region in Fig.~\ref{fig:radmass} can always be avoided for both compact stars. 

\begin{figure}[t]
\centering
\includegraphics[width=.9\linewidth]{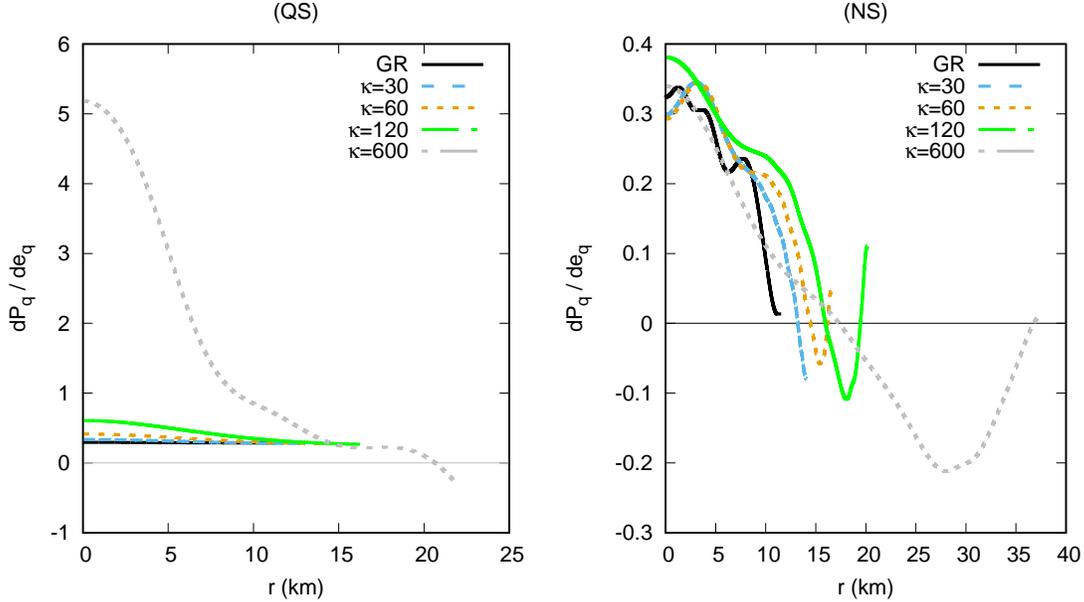}
\caption{The apparent sound speed for NS in the right panel, and for QS in the left panel.}
\label{fig:sound}
\end{figure}

The maximum masses of NS and QS as a function of $\kappa$ are shown in Fig.~\ref{fig:MaxMasses}. To constrain the lower limit of $\kappa$ we use the maximum mass constraints which are taken from Refs.\cite{Compact1,Compact2,Rezzolla:2017aly}. Note that constraints from  Refs.\cite{Compact1,Compact2} are used in Ref.~\cite{QISR2016} to obtain the lower limit of $\kappa$ in their work. They used also BSP parameter set as representative of acceptable EOS within RMF model.  It is obvious from  Fig.~\ref{fig:MaxMasses} in the case of NS, if we use stiffer EOS than the one of BSP, the lower limit of $\kappa$ becomes smaller while if we use softer EOS than that of BSP it becomes larger. Furthermore, if we use extreme stiff EOS for example by using NL3 parameter set but with the hyperon contributions are excluded, we can obtain lower limit of $\kappa <$ 0. Similar situation occurs also in the case of QS. Here for NS if we use constraint from Antoniadis {\it et. al}~\cite{Compact2}, the lower limit of $\kappa$ is  $\kappa \ge$ 6 ($\kappa_g \ge$ 0.01). This value is quite compatible with the one obtained in Ref.~\cite{QISR2016}. However, if we use most recent gravitational wave constraint from Rezzola {\it et. al}~\cite{Rezzolla:2017aly} then we obtain larger lower limit, i.e., $\kappa \ge$ 16 ($\kappa_g \ge$ 0.027). While for QS with  QSK046,  if we use constraint from Antoniadis {\it et. al}~\cite{Compact2}, the lower limit of $\kappa$ is  $\kappa \ge$ 1.4 ($\kappa_g \ge$ 0.002) while if we use the constraint from Rezzola {\it et. al} ~\cite{Rezzolla:2017aly} yields $\kappa \ge$ 9.5 ($\kappa_g \ge$ 0.016). In addition it is shown in the middle panel of Fig.~\ref{fig:MaxMasses}  that QS  within CIDDM with  vector Coulomb term EOS can also achieve higher maximum mass with $\kappa$ value depends on the stiffness of the corresponding EOS. It means that the vector-model of QS can also achieve maximum mass higher than 2.0 M$_\odot$. This maximum mass cannot be achieved if one uses standard GR in the QS calculation within the CIDDM with vector Coulomb term in the EOS \cite{QS2016}.

The radius of canonical NS and QS as a function of $\kappa$ are shown in Fig.~\ref{fig:1.4MSunradii}. For comparison, we compare the results with observation results taken from Refs.\cite{Steiner:2017vmg,Steiner:2015aea,Nattila:2015jra,Guillot:2014lla,Annala:2017llu}. As mentioned previously that different to those of compact star masses, the measured radii of compact stars hampered by uncertainty due to systematics. Therefore, for example in canonical mass of compact stars, the overall range of the radius in average becomes quite wide. It is obvious in Fig.~\ref{fig:1.4MSunradii} that the allowed $\kappa$ depends sensitively on the specific radius canonical mass range used. It means that the uncertainty of the systematic in compact stars radii measurements, specially in compact stars with canonical mass can affect the $\kappa$ range obtained. It can be seen that for the case of QS the range depends on the stiffness of the EOS used. It is happen not only for the one with scalar Coulomb but also for the one with vector Coulomb. The compatibility of QS canonical mass radius to the recent observations data is also quite well. On the other hand for NS, the stiffness of EOS used do not affect significantly the predicted allow $\kappa$ and the  compatibility of QS canonical mass radius to the recent radii observations data is only happen on the constraints which have a quite large radius limit, i.e., $R_{1.4 M_{\odot}}\approx$ 14 m. For example, if we use the constraint from Analla {\it et. al}~\cite{Annala:2017llu} or Steiner  {\it et. al}~\cite{Steiner:2017vmg} the predicted $\kappa$ by using BSP or G2 EOSs is  $\kappa\approx$ 0.45  ($\kappa_g \approx$ 0.0006). This result is significantly smaller compared the ones obtained from maximum mass and compactness constraints. The reason is due to the fact that NS has dilute crust EOS while the QS has not. Therefore, for low mass NS the radius is relative large. It means the uncertainty in NS crust EOS used also plays quite a role to add uncertainty in the canonical mass radius or the radii of other small mass constraints.

In the following we will discuss the compatibility of the apparent EOS of NS and QS within the EiBI theory with constraint from energy conditions. The simultaneous fulfillment of the energy conditions of Eqs.~(\ref{nec}-\ref{cec}) can be synthesized and observed from the relations: $\epsilon \geq 0$, and $\frac{P}{\epsilon} \geq - \frac{1}{3}$, with ${|P|} \leq {\epsilon}$. Furthermore, the fulfillment of corresponding energy conditions for the apparent EOSs of realistic models of NS and QS  can be observed clearly from the profile of maximum masses shown in Fig.~\ref{fig:NS} for NSs and Fig.~\ref{fig:QS} for QSs. It can be seen in the left-lower panel of Fig.~\ref{fig:NS} and Fig.~\ref{fig:QS} that the apparent energy densities of NSs and QSs still have positive value everywhere ($\epsilon_q \geq 0$). The right-upper panel of Fig.~\ref{fig:NS} and Fig.~\ref{fig:QS} show the ratio value of apparent pressure to apparent energy condition $\frac{P_q}{\epsilon_q}$.  We expect  by observing the change of trend due to the varying $\kappa$ value in upper right panel of Fig.~\ref{fig:NS} and Fig.~\ref{fig:QS}, that at a very large $\kappa$ value the apparent EOSs  will eventually violate the tighter SEC, namely $\frac{P_q}{\epsilon_q}\geq - \frac{1}{3}$. However, even up to the value of $\kappa = 600 $ ( $\kappa_9 = 1 $) the  corresponding constraint is still not saturated. It means that if our expectation is right, this energy condition will be violated only at the extremely large $\kappa$. In the upper panels of Fig.~\ref{fig:NS}, one can observe that the apparent pressure of NSs becomes negative in the region close to the surface (crust) of the stars. The effect becomes larger if higher $\kappa$ is used in the calculation. Similar situation happens in QSs (upper panels of Fig.~\ref{fig:QS}), where the apparent pressure becomes negative in the region close to the surface of the stars. It is interesting to observe that the apparent pressure of QS is not going to be zero at the surface edge, while the NS goes back to zero there. This happens because NSs have a dilute crust while QSs do not. Therefore, the effect of effective contribution of the gravity within EiBI in matter becomes dominant in region close to NS surface. However, if we compare the apparent EOS profile  of maximum mass of NS with the one of QS at a very large $\kappa$, the behavior is significantly different. The apparent pressure of NS at $\kappa=600$ ($\kappa_g$= 1.0), for example, becomes negative from the center of NS until the surface, while the negative apparent pressure in the center starts to appear at $\kappa\simeq 460$ ($\kappa_g\simeq$ 0.77). On the other hand, in QS the negative apparent pressure appears only on the surface. Note that the actual EOS coincide with the apparent EOS in the case of GR (black line) and the actual EOS satisfies all of the energy conditions. Furthermore, the actual EOS has only positive pressure value everywhere.  Therefore, from the results in Fig.~\ref{fig:NS} and Fig.~\ref{fig:QS}, we can conclude that up to quite a wide range of $\kappa$ the apparent EOS for NSs and QSs satisfy all of the energy conditions simultaneously. On the other hand, if we observe that for NS with  $\kappa \le 47$ ($\kappa_g\le$ 0.079) at the surface the maximum $ |P_q| \lesssim$  1 $\rm MeV ~fm^{-3}$, while for QS with $\kappa \le 13$ ($\kappa_g\le$ 0.022) at the surface the maximum $ |P_q|$ is relative small i.e., more or less at the same order of magnitude with the one of NS. We can expect that for NSs with $\kappa \le 47$ ($\kappa_g\le$ 0.079) and QSs with  $\kappa \le 13$ ($\kappa_g\le$ 0.022) within EiBI, the corresponding EOSs do not significantly violate the general physical EOS requirement for acceptable interior solution for static fluid spheres of GR i.e., the requirement that $P_q$ should be positive inside the star. Unfortunately it is not really the case. We show the profiles of square of the apparent sound of speed for NS and QS maximum masses in Fig.~\ref{fig:sound} to see the impact of the appearance of negative value $P_q$ in the region close to surface. It can be seen in the left panel that in the QS case, the  apparent squared sound of speed is still obey the requirement for  $\kappa \le 120$ ($\kappa_g\approx$ 0.20), while for the case of NS (as shown in right panel) such value in the region close to the surface becomes negative if  $\kappa > 0$ and the corresponding absolute values becomes larger when $\kappa$ increases. Once again, the reason of this is the effect of effective contribution of the gravity within EiBI in matter becomes significant in region close to NS surface which generate such effect in apparent sound of speed. Note that for GR, the square of apparent sound of speed of NS or QS are always positive everywhere because $P_q=P$. It means this effect is independent on the compact star EOSs used.

\section{Conclusion}
\label{sec_conclu}

In this work, the compactness, maximum masses and radii of canonical mass constraints of NSs and QSs are employed to investigate the upper and lower limits of $\kappa$ parameter in the EiBI theory. We also investigate the compatibility of the apparent EOSs of NS and QS with energy conditions as well as general physical requirement of static fluid spheres. 

For NS, the EOS of the star core is calculated using the RMF model with BSP parameter set~\cite{SB2012} under which the standard SU(6) prescription and hyperon potential depths~\cite{JSB_AG} are utilized to determine the hyperon coupling constants, for inner crust we use polytropic EOS while for the outer crust EOS the one proposed by R\"uster {\it et al.}~\cite{Ruester:2005fm}  is used. To check the dependence of  $\kappa$ range obtained with the stiffness of the EOS, we use also EOSs calculated by using NL3 parameter set as a representation of stiff EOS and G2 parameter set as a representation of soft EOS within RMF model. For QS we used EOS based on the CIDDM model with additional scalar Coulomb term of strange quark matter i.e., QSK046 parameter set. Similar to NS, for QS we also check the dependence of  $\kappa$ range obtained with the stiffness of the EOS for scalar and vector Coulomb cases. We have found that for large $\kappa$ value, both NS and QS do not exceed causality restrictions because the compactness of NS and QS is saturated after passing a certain large $\kappa$ value. This finding is in agreement with the result obtained in Ref. \cite{Delsate12}. The variation of $\kappa$ causes the CIDDM model with vector-Coulomb of QS can reach the maximum mass $\ge2M_\odot$. This happens because a large value of $\kappa$ can yield a large maximum mass value of QS compared to that of GR. This results complement the results found in Ref.\cite{QS2016}. From compactness~\cite{Compact3} and most recent maximum masses from gravitational wave~\cite{Rezzolla:2017aly} constraints analysis we conclude that the upper limit of $\kappa$ value of NS+hyperons predicted by BSP parameter set is  $16 \le \kappa \le 47$ ($ 0.027 \le\kappa_g\le 0.079$). This results complement the constraints of $\kappa$ value of NS result obtained by the authors in Refs.~\cite{Avelino12,PCD2011,QISR2016}. Furthermore, we also obtain that the range of $\kappa$ value of QS within CIDDM model with additional scalar Coulomb term is 
$9.5 \le \kappa \le 13$ ($0.016 \le \kappa_g\le 0.022$). We have found also thatthese constrant ranges depend significantly on the the EOS employed. If the NS or QS EOS becomes stiffer, the upper limit of $\kappa$ from compactness constraint becomes smaller and lower limit of  $\kappa$ from maximum masses constraint becomes larger.  We also observe that the uncertainty of the systematic in canonical mass compact stars radii measurements data can affect the $\kappa$ range. 

The apparent EOS of NSs and QSs with wide range of $\kappa$ can satisfy the energy conditions constraints. It is indicated that for extreme large value of $\kappa$, the strong energy condition can be violated. The fraction of $\frac{P_q}{\epsilon_q}$ will be less than $- \frac{1}{3}$ in such extremely large of value of $\kappa$. The general physical EOS requirement for acceptable interior solution for static fluid spheres of GR can be also violated by apparent EOS. In the NSs case, the square of the speed of sound of the corresponding apparent EOSs in the region near the NS surface can be negative due to the dominant role of negative pressure from effective gravity contribution within EIBI in NS surface. However, this is not make problem with EiBI theory because the apparent EOS which is defined in $q$ metric, is not a physical quantity.

%%%%%%%%%%%%%%%%%%%%%%%%%%%%%%%%%%%%%%%%%%%%%%%%%%%%%%%%%%%%%%%%%%%%%%%%%%
\section*{ACKNOWLEDGMENT}
We thank Ilham Prasetyo for useful discussions during the course of this work.
AS is partially supported by the UI's PITTA grant No. 622/UN2.R3.1/ HKP.05.00/2017. HSR is partially supported by the UI's PITTA grant No. 656/UN2.R3.1/HKP.05.00/2017.

%%%%%%%%%%%%%%%%%%%%%%%%%%%%%%%%%%%%%%%%%%%%%%%%%%%%%%%%%%%%%%%%%%%%%%%%%

\end{document}